\documentclass{article}
\usepackage{amsmath}
\usepackage{tikz}
\usepackage{changes}

\title{Quantum cipher based on phase inversion} 

\author{Vikram Menon$^{1}$ \& Ayan Chattopadhyay$^{2\ast}$}

\begin{document} 

\maketitle 



\begin{abstract}
  We present a quantum version of a cipher used in cryptography where the
  message to be communicated is encoded into the relative phase of a quantum state
  using the shared key. The encoded quantum information carrying the message is
  actually sent to the recepient over a quantum channel, assuming the proper
  secrecy of shared key between peers. 
\end{abstract}


\section{Introduction}
The one-time pad, also known as the Vernam Cipher, invented and
patented\cite{verpat,perfcpr} by Gilbert Vernam in 1917, is a
perfect cipher.\cite{verpat,perfcpr,otp} The core idea is that any message
can be transformed into any cipher (of the same length) by a pad, such that all
transformations are equally likely. The one-time pad encryption scheme is
provably unbreakable if (i) the key is at least the size of the plaintext,
(ii) the key is truly random, and(iii) the key is used only once. The requirement
of only one time key usage makes the one-time pad impractical when the amount of
date to be securely communicated is huge.

The more practical ciphers are the block ciphers, like DES, 3DES, AES\cite{aes},
which operate on a block of message depending on the key size.
A given key is reused for mutiple message blocks. This makes the cipher prone to various kinds
of attacks, like the cloning attack, cryptanalysis attack, chosen ciphertext/plaintext attack etc.
The security of these ciphers depend mainly upon the strength of the key
(randomness) and the algorithm used for encryption and decryption in terms of
{\em confusion} and {\em diffusion} created.

In 1984, Charles H. Bennett and Gilles Brassard described the first completely
secure quantum key distribution algorithm often known as the BB84 algorithm.\cite{qkd}
And in 1997, Peter Shor\cite{qfac} provided a quantum algorithm that can break
the widely used RSA system, using quantum computers, with remarkable ease.
Together the developments showed that a completely secure, efficient, and fast
means of delivering confidential information is achievable using the laws of
quantum mechanics and quantum computers.
Ciphers based on quantum states, therefore, would be more appropriate  because
of the property that an unknown state cannot be copied. Hence, none of the
attacks in classical cryptography would be applicable here. In recent times
there are many attempts and constructions for quantum ciphers. One such
attempt was made by Abdullha, A. A. et. al.\cite{qrng}, where they used quantum
random number generator and half adder for encryption and transmit with the
help of BB84 protocol.

In this paper, a quantum version of the cipher is proposed, which
utilizes the benefit of superior delivery efficiency provided by modern
telecommunication, and snoop-detection capability of the BB84 algorithm.
The quantum version exploits the fact that (1) an unknown quantum state
cannot be cloned and (2) its relative phase cannot be measured. The message
to be sent is encoded into a quantum state by altering the relative phase
using a pre-established shared key, via BB84 or any other quantum key
distribution protocol. The information transmitted is a quantum
superposition state with uniform probability and relative phase distribution.

\section{The algorithm}
The proposed algorithm requires a pre-established shared key between the
communicating parties, which can be achieved by the BB84 or similar QKD
(Quantum Key Distribution)
protocol. It is assumed that QKD is secured enough and it would be hard
to know any information about key.

The components of the proposed cryptosystem (encoder and decoder)
\begin{enumerate}
  \item Hadamard transformation.
  \item Oracle, $\Lambda_{k}$ defined in section 1.2, responsible for key state phase inversion.
  \item Another oracle, $\Upsilon_{k,d}$ defined in section 1.3, for multiple phase inversion.
\end{enumerate}
are captured in the following Figure 1.\\

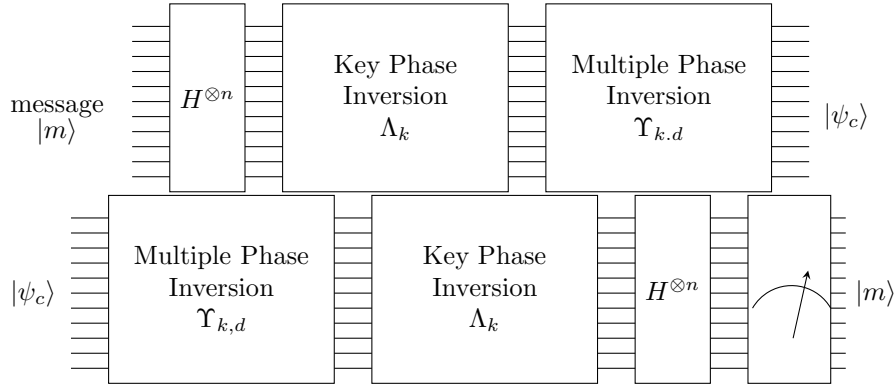
\begin{tikzpicture}
  \draw (0,1.1) node {message};
  \draw (0,.8) node {$|m\rangle$};
  
  \draw (1,2.2) -- (1.5,2.2);
  \draw (1,2.0) -- (1.5,2.0);
  \draw (1,1.8) -- (1.5,1.8);
  \draw (1,1.6) -- (1.5,1.6);
  \draw (1,1.4) -- (1.5,1.4);
  \draw (1,1.2) -- (1.5,1.2);
  \draw (1,1.0) -- (1.5,1.0);
  \draw (1,.8) -- (1.5,.8);
  \draw (1,.6) -- (1.5,.6);
  \draw (1,.4) -- (1.5,.4);
  \draw (1,.2) -- (1.5,.2);

  \draw (1.5,0) rectangle (2.5,2.5) node [midway=center] {$H^{\otimes n}$};

  \draw (2.5,2.2) -- (3,2.2);
  \draw (2.5,2.0) -- (3,2.0);
  \draw (2.5,1.8) -- (3,1.8);
  \draw (2.5,1.6) -- (3,1.6);
  \draw (2.5,1.4) -- (3,1.4);
  \draw (2.5,1.2) -- (3,1.2);
  \draw (2.5,1.0) -- (3,1.0);
  \draw (2.5,.8) -- (3,.8);
  \draw (2.5,.6) -- (3,.6);
  \draw (2.5,.4) -- (3,.4);
  \draw (2.5,.2) -- (3,.2);

  \draw (3,0) rectangle (6,2.5) node [midway=center] {\begin{tabular}{c} Key Phase \\ Inversion \\ $\Lambda_{k}$\end{tabular}};

  \draw (6,2.2) -- (6.5,2.2);
  \draw (6,2.0) -- (6.5,2.0);
  \draw (6,1.8) -- (6.5,1.8);
  \draw (6,1.6) -- (6.5,1.6);
  \draw (6,1.4) -- (6.5,1.4);
  \draw (6,1.2) -- (6.5,1.2);
  \draw (6,1.0) -- (6.5,1.0);
  \draw (6,.8) -- (6.5,.8);
  \draw (6,.6) -- (6.5,.6);
  \draw (6,.4) -- (6.5,.4);
  \draw (6,.2) -- (6.5,.2);

  \draw (6.5,0) rectangle (9.5,2.5) node [midway=center] {\begin{tabular}{c} Multiple Phase \\ Inversion \\ $\Upsilon_{k.d}$\end{tabular}};

  \draw (9.5,2.2) -- (10,2.2);
  \draw (9.5,2.0) -- (10,2.0);
  \draw (9.5,1.8) -- (10,1.8);
  \draw (9.5,1.6) -- (10,1.6);
  \draw (9.5,1.4) -- (10,1.4);
  \draw (9.5,1.2) -- (10,1.2);
  \draw (9.5,1.0) -- (10,1.0);
  \draw (9.5,.8) -- (10,.8);
  \draw (9.5,.6) -- (10,.6);
  \draw (9.5,.4) -- (10,.4);
  \draw (9.5,.2) -- (10,.2);

  \draw (10.5,1.0) node {$|\psi_{c}\rangle$};

\end{tikzpicture}

\begin{tikzpicture}
  \draw (0,1.2) node {$|\psi_{c}\rangle$};
  
  \draw (.5,2.2) -- (1,2.2);
  \draw (.5,2.0) -- (1,2.0);
  \draw (.5,1.8) -- (1,1.8);
  \draw (.5,1.6) -- (1,1.6);
  \draw (.5,1.4) -- (1,1.4);
  \draw (.5,1.2) -- (1,1.2);
  \draw (.5,1.0) -- (1,1.0);
  \draw (.5,.8) -- (1,.8);
  \draw (.5,.6) -- (1,.6);
  \draw (.5,.4) -- (1,.4);
  \draw (.5,.2) -- (1,.2);

  \draw (1,0) rectangle (4,2.5) node [midway=center] {\begin{tabular}{c} Multiple Phase \\ Inversion \\ $\Upsilon_{k,d}$\end{tabular}};
  \draw (4,2.2) -- (4.5,2.2);
  \draw (4,2.0) -- (4.5,2.0);
  \draw (4,1.8) -- (4.5,1.8);
  \draw (4,1.6) -- (4.5,1.6);
  \draw (4,1.4) -- (4.5,1.4);
  \draw (4,1.2) -- (4.5,1.2);
  \draw (4,1.0) -- (4.5,1.0);
  \draw (4,.8) -- (4.5,.8);
  \draw (4,.6) -- (4.5,.6);
  \draw (4,.4) -- (4.5,.4);
  \draw (4,.2) -- (4.5,.2);

  \draw (4.5,0) rectangle (7.5,2.5) node [midway=center] {\begin{tabular}{c} Key Phase \\ Inversion \\ $\Lambda_{k}$\end{tabular}};

  \draw (7.5,2.2) -- (8,2.2);
  \draw (7.5,2.0) -- (8,2.0);
  \draw (7.5,1.8) -- (8,1.8);
  \draw (7.5,1.6) -- (8,1.6);
  \draw (7.5,1.4) -- (8,1.4);
  \draw (7.5,1.2) -- (8,1.2);
  \draw (7.5,1.0) -- (8,1.0);
  \draw (7.5,.8) -- (8,.8);
  \draw (7.5,.6) -- (8,.6);
  \draw (7.5,.4) -- (8,.4);
  \draw (7.5,.2) -- (8,.2);
  
  \draw (8,0) rectangle (9,2.5) node [midway=center] {$H^{\otimes n}$};

  \draw (9,2.2) -- (9.5,2.2);
  \draw (9,2.0) -- (9.5,2.0);
  \draw (9,1.8) -- (9.5,1.8);
  \draw (9,1.6) -- (9.5,1.6);
  \draw (9,1.4) -- (9.5,1.4);
  \draw (9,1.2) -- (9.5,1.2);
  \draw (9,1.0) -- (9.5,1.0);
  \draw (9,.8) -- (9.5,.8);
  \draw (9,.6) -- (9.5,.6);
  \draw (9,.4) -- (9.5,.4);
  \draw (9,.2) -- (9.5,.2);

  \draw (9.5,0) rectangle (10.6,2.5);
  \draw (10.6,1) arc (30:150:.6cm);
  \draw [->,>=stealth] (10.1,.6) -- (10.3,1.5);

  \draw (10.6,2.2) -- (10.8,2.2);
  \draw (10.6,2.0) -- (10.8,2.0);
  \draw (10.6,1.8) -- (10.8,1.8);
  \draw (10.6,1.6) -- (10.8,1.6);
  \draw (10.6,1.4) -- (10.8,1.4);
  \draw (10.6,1.2) -- (10.8,1.2);
  \draw (10.6,1.0) -- (10.8,1.0);
  \draw (10.6,.8) -- (10.8,.8);
  \draw (10.6,.6) -- (10.8,.6);
  \draw (10.6,.4) -- (10.8,.4);
  \draw (10.6,.2) -- (10.8,.2);

  \draw (11.2,1.2) node {$|m\rangle$};
  
\end{tikzpicture}

The message $|m\rangle$ to be sent is first passed through a Hadamard transformation
to create an equal superposition state. Next, the phase inversion is applied to
invert the key state phase. Finally the multiple phase inversion transformation
(another Oracle) is applied to invert the phases of exactly half of the basis states.
The outcome of this is an encrypted quantum state $|\psi_{c}\rangle$ with a uniform
probability and relative phase distributions. The transmitted quantum state act as a
carrier for the message, analogous to the FM transmission, where the audio wave to be
transmitted is encoded in the frequency of a high frequency carrier wave.

The recipient can apply the transformation to $|\psi_{c}\rangle$ in reverse, the multiple
phase inversion followed by the key phase inversion and finally the Hadamard
transformation, to retrieve the original message $|m\rangle$.

\subsection{Hadamard transformation}
The Hadamard transformation, irrespective of the input, creates an equal
superposition state, i.e. a uniform distribution of all possible n-qubit states
of the message space, say {\em M}, of size $N = 2^{n}$.

\begin{equation}
|\psi_{m}\rangle = H^{\otimes n} (|m\rangle) = \frac{1}{\sqrt{N}} \sum_{x=0}^{N-1} (-1)^{m.x}|x\rangle
\end{equation}

\subsection{Key phase inversion}
The phase inversion operator, ($\Lambda_{k}$), acts as an oracle and is defined as
\begin{equation}
  \Lambda_{k} = I - 2| k\rangle\langle k|
\end{equation}
where $|k\rangle$ is the quantum key state derived from the shared key $k$ and
$I$ the identity operator.

The application of this operator on the input state $|\psi_{m}\rangle$, equation (1), 
creates a coupling between the input state and the key state.
\begin{align}
  |\psi_{c}^{'}\rangle = \Lambda_{k} |\psi_{m}\rangle & = I|\psi_{m}\rangle - 2\frac{1}{\sqrt{N}} \sum_{x=0}^{N-1} (-1)^{m.x} |k\rangle\langle k|x\rangle \nonumber \\
         & = |\psi_{m}\rangle - \frac{2}{\sqrt{N}} (-1)^{m.k}|k\rangle
\end{align}

This marking of the key state accomplishes the encoding.

The application of the same operator retrieves the input state $|\psi_{m}\rangle$, 
\begin{align}
  \Lambda_k |\psi_{c}^{'}\rangle &= (I - 2 |k\rangle\langle k|)(|\psi_{m}\rangle - \frac{2}{\sqrt{N}} (-1)^{m.k}|k\rangle)\nonumber \\
  &= |\psi_{m}\rangle - \frac{2}{\sqrt{N}} (-1)^{m.k}|k\rangle - \frac{2}{\sqrt{N}} (-1)^{m.k} |k\rangle + \frac{4}{\sqrt{N}} (-1)^{m.k} |k\rangle \nonumber \\
  &= |\psi_{m}\rangle
\end{align}

\subsection{Multiple phase inversion}
The multiple phase inversion transformation, say $\Upsilon_{k,d}$, performs phase
inversion of multiple basis states as follows.

\begin{enumerate}
\item Select one $r$, where $r|N$ ($r$ divides $N$) and $r<\frac{N}{2}$. Now say,
  $d = \frac{N}{r}$, where $N = 2^{n}$ for $n$ qubits.
  
\item Start with the key $k$ position of the state $|\psi_{c}^{'}\rangle$ and invert the
  phases of next $d$ consecutive states, i.e. states $|(k + pd) \mod N\rangle$ to
  $|(k + (p+1)d - 1) \mod N\rangle$. Skip the next $d$ states from
  $|(k + (p+1)d) \mod N\rangle$ to $|(k + (p+2)d - 1) \mod N\rangle$. Here
  $0 \le p \le \frac{N}{2d}$ and $p \in {N \cup {0}}$, here ${N}$ is set of natural
    numbers.
\end{enumerate}

The transformation can be defined as,
\begin{equation}
  |\psi_{c}\rangle = \Upsilon_{k,d} |\psi_{c}^{'}\rangle
\end{equation}

The distribution of inverted vs. non-inverted phase states ($|\psi_{c}\rangle$,	 being the complete superposed states) will vary
on each unique choice of key state and $d$. Hence, the guessing of the state distribution is not
possible in this construction. It can be easily visualized that with this algorithm,
phases of half of the basis states will be inverted. $'d'$ can be uniquely defined
for a given key and can be arrived at as part of the key exchange process.

\subsection{Security}
The key state phase inversion operator, $\Lambda_k$ defined by equation (2), when applied to the state
$|\psi_{m}\rangle$, resulted in the inversion of the key state $|k\rangle$ as given by equation (3).
It can be rewritten as follows,
\begin{align}
  |\psi_{c}^{'}\rangle &= \Lambda_{k} |\psi_{m}\rangle \nonumber \\
         & = |\psi_{m}\rangle - \frac{2}{\sqrt{N}} (-1)^{m.k}|k\rangle \nonumber \\
         & = |\psi_{m-k}\rangle - \frac{1}{\sqrt{N}} (-1)^{m.k}|k\rangle
\end{align}
where $|\psi_{m-k}\rangle = |\psi_{m}\rangle - \frac{1}{\sqrt{N}}|k\rangle$ is the superposition
of all the basis states, except the key state.

The probability of the phase inverted key state be $P_{k} = \frac{1}{N}$ and each of
the remaining $(N-1)$ non-inverted states $(\forall x\in |x\rangle, k\notin x)$ be $P_{x} = \frac{1}{N}$.
The ratio of the probability of inverted and non-inverted states 
\begin{equation}
\frac{P_{k}}{P_{x}} = \frac{N}{N} = 1
\end{equation}
is probabilistically indistinguishable in this case.

The adversary can only see the transmitted state as given equation {3}, he/she has the power to
apply Hadamard transform on the transmitted state. However, since adversary does not know the
shared key state $|k\rangle$, $\Lambda_k$ remains private, applying Hadamard transform cannot
return back the original message state $|\psi_{m}\rangle$. The adversary can only guess the
construction of $\Lambda_k$ in $O(\sqrt{N})$ running time.

Alternatively, adversary can set up chosen plaintext attack by apply inversion against mean
operator (Ref: Grover's inversion against mean) to check if it could leak some information,
or can guess any inherent biasness.

Let us introduce the inversion against mean operator defind as
$\mu_{m} = (2|\psi_{m}\rangle\langle \psi_{m}| - I)$, where $|\psi_{m}\rangle$ is total state.
The inner product of key state and total state is given by $\psi_{m}|k\rangle = \langle k|\psi_{m}\rangle = \frac{1}{\sqrt{N}}$.
The application of $\mu_{m}$ to $|\psi_{c}^{'}\rangle$ the will result in
\begin{align}
  |\phi_{m}\rangle = \mu_{m} |\psi_{c}^{'}\rangle &= (2|\psi_{m}\rangle\langle \psi_{m}| - I)(|\psi_{m}\rangle - \frac{2}{\sqrt{N}}|k\rangle) \nonumber \\
  &= 2|\psi_{m}\rangle\langle \psi_{m}|\psi_{m}\rangle - 2\frac{2}{\sqrt{N}}|\psi_{m}\rangle\langle \psi_{m}|k\rangle - (|\psi_{m}\rangle + \frac{2}{\sqrt{N}}|k\rangle) \nonumber \\
  &= (1 - \frac{4}{N}) |\psi_{m}\rangle + \frac{2}{\sqrt{N}} |k\rangle
\end{align}

Let us choose one use case having message having all 0 i.e. $|\psi_{0}\rangle = H^{\otimes n} (|00...0\rangle)$.
Equation 8 can then be rewritten as:
\begin{align}
  |\phi_{0}\rangle = (1 - \frac{4}{N}) |\psi_{0}\rangle + \frac{2}{\sqrt{N}} |k\rangle
\end{align} 
 
To extract out non-inverted phase states, equation 6 is used in equation 9 and we get
\begin{align}
  |\phi\rangle &= (1 - \frac{4}{N}) |\psi_{0}\rangle + \frac{2}{\sqrt{N}} |k\rangle \nonumber \\
  &= (1 - \frac{4}{N}) |\psi_{0-k}\rangle + \frac{(1 - \frac{4}{N})}{\sqrt{N}}|k\rangle + \frac{2}{\sqrt{N}} |k\rangle \nonumber \\
  &= (1 - \frac{4}{N}) |\psi_{0-k}\rangle + \frac{3N - 4}{N\sqrt{N}}|k\rangle  
\end{align}

The probability of the phase inverted state is therefore given by $P_{k} = (\frac{3N - 4}{N\sqrt{N}})^{2}$
and each of the remaining $(N-1)$ non-inverted states ($\forall x\in |x\rangle , k\notin x$) by
$P_{x} = \frac{(1 - \frac{4}{N})^{2}}{N}$. Looking for the same probability ratio yields
\begin{align}
  \frac{P_{k}}{P_{x}} &= \frac{(\frac{3N - 4}{N\sqrt{N}})^{2}}{\frac{(1 - \frac{4}{N})^{2}}{N}} \nonumber \\
  &= (\frac{3N - 1}{N - 4})^{2} \nonumber \\
  &= (\frac{3 - \frac{4}{N}}{1 - \frac{4}{N}})^{2}
\end{align}
which, for large $N$, will reduce to $\lim \frac{P_{k}}{P_{x}} \rightarrow 9$.

It is noticed that there is a biasness of probability distribution between phase inverted
state and non-inverted state. The adversary has the freedom to apply $\mu_{0}$ again on the
output of first $\mu_{0}$ operation, if doing so, it will be observed from the below result
that the encrypted transmitted state would be emerged.
\begin{align}
  |\phi\rangle &= (2|\psi_{0}\rangle\langle \psi_{0}| - I)(1 - \frac{4}{N}) |\psi_{0}\rangle + \frac{2}{\sqrt{N}} |k\rangle \nonumber \\
  &= (1 - \frac{4}{N}) |\psi_{0}\rangle + \frac{4}{N}|\psi_{0}\rangle - \frac{2}{\sqrt{N}} |k\rangle \nonumber \\
  &= |\psi_{0}\rangle - \frac{2}{\sqrt{N}} |k\rangle
\end{align}

So, with the repeated use of $\mu_{0}$, would emerge the above alternative pattern (alternative
repeation of eqution 8 equation 12). Though, the adversary cannot be able to make out any
useful infomarion but it shows a little biasness in probability distribution of the inverted
key states over rest of the individual (non-inverted) message states. This violates the
Shannon's secrecry clause for encryption\cite{shan}.

In order to solve the biasness problem, we could phase invert $M$ states and $M > 1$. The
objective is to show that if $M = \frac{N}{2}$ the biasness can be eliminated and we could
show the inverted phase states and the rest of the states will be indistinguishable.

Multiple phase inversion is actually a chain of single phase inversion of $M$ times with
the corresponding phase inversion states are $|k_{1}\rangle, |k_{2}\rangle, ... |k_{m}\rangle$,
these states remain private between the communicating parties.

\begin{align}
&(I - 2| k_{1}\rangle\langle k_{1}|) |\psi_{m}\rangle = |\psi_{m}\rangle - \frac{2}{\sqrt{N}} (-1)^{m.k_{1}}|k_{1}\rangle \nonumber \\
&(I - 2| k_{2}\rangle\langle k_{2}|) (|\psi_{m}\rangle - \frac{2}{\sqrt{N}} (-1)^{m.k_{1}}|k_{1}\rangle)
  = |\psi_{m}\rangle - \frac{2}{\sqrt{N}} (-1)^{m.k_{1}}|k_{1}\rangle - \frac{2}{\sqrt{N}} (-1)^{m.k_{2}}|k_{2}\rangle \nonumber \\
& ... \nonumber
\end{align}
similarly goes on up to $M$ states and the final would look like 
\begin{align}
  (I - 2| k_{m}\rangle\langle k_{m}|) (|\psi_{m}\rangle &- \frac{2}{\sqrt{N}} (-1)^{m.k_{1}}|k_{1}\rangle \nonumber\\
  &- \frac{2}{\sqrt{N}} (-1)^{m.k_{2}}|k_{2}\rangle \nonumber\\
  &... \nonumber \\
  &- \frac{2}{\sqrt{N}} (-1)^{m.k_{m-1}}|k_{m-1}\rangle) \nonumber \\
  &= |\psi_{m}\rangle - \frac{2}{\sqrt{N}} (-1)^{m.k_{1}}|k_{1}\rangle - \frac{2}{\sqrt{N}} (-1)^{m.k_{2}}|k_{2}\rangle \nonumber\\
  &... - \frac{2}{\sqrt{N}} (-1)^{m.k_{m}}|k_{m}\rangle
\end{align}
Without any loss of generality, this can be expressed as
\begin{align}
  |\phi\rangle = \Lambda_k|\psi_{m}\rangle &= |\psi_{m}\rangle - \frac{2}{\sqrt{N}} \sum_{i=0}^{M-1} (-1)^{m.k_{i}}|k_{i}\rangle \nonumber\\
  &= |\psi_{m-k}\rangle - \frac{1}{\sqrt{N}} \sum_{i=0}^{M-1} (-1)^{m.k_{i}}|k_{i}\rangle
\end{align}

Now, the probability of the $M$ consolidated phase inverted key states is $P_{k} = M.(\frac{1}{\sqrt{N}})^{2}$ and
$(N-M)$ non-inverted states $P_{x} = (N - M).(\frac{1}{\sqrt{N}})^{2}$ and the ratio 
\begin{equation}
\frac{P_{k}}{P_{x}} = \frac{M}{(N - M)}
\end{equation}
which reduces to $\frac{P_{k}}{P_{x}} = 1$ when $M = \frac{N}{2}$.
 
Thus proved that there is no biasness in probability distribution between $M$ phase inverted
states with the rest of the non-inverted states. Since adversary does not know the shared key
state $|k\rangle$, $\Lambda_k$ remains private, applying Hadamard transform on the transmitted
message will not reveal any information. The adversary, however can chose to apply inversion
against mean operator (Ref: Grover's inversion against mean) to launch 'chosen plaintext attack',
taking the similar argumental approach as used during single phase inversion analysis (changing
notation of $\psi_{m}$ to $\psi_{0}$), 

\begin{align}
  |\phi\rangle = \mu_{0} |\psi_{0}\rangle &= (2|\psi_{0}\rangle\langle \psi_{0}| - I)(|\psi_{0}\rangle - \frac{2}{\sqrt{N}} \sum_{i=0}^{M-1} (-1)^{m.k_{i}}|k_{i}\rangle \nonumber \\
  &= 2|\psi_{0}\rangle\langle \psi_{0}|\psi_{0}\rangle - 2\frac{2}{\sqrt{N}}\sum_{i=0}^{M-1} (-1)^{m.k_{i}}|\psi_{0}\rangle\langle \psi_{0}|k_{i}\rangle - (|\psi_{0}\rangle + \frac{2M}{\sqrt{N}}|k\rangle) \nonumber \\
  &= (1 - \frac{4M}{N}) |\psi_{0}\rangle + \frac{2}{\sqrt{N}} \sum_{i=0}^{M-1} (-1)^{m.k_{i}}|k_{i}\rangle
\end{align}

Since $|\psi_{0}\rangle = \frac{N - M}{\sqrt{N}}|\psi_{0-k}\rangle + \frac{M}{\sqrt{N}}|k\rangle$, 
to extract inverted and non-inverted phase states, equation 16 can be re-written as:
\begin{align}
  |\phi\rangle &= (1 - \frac{4M}{N}) |\psi_{0}\rangle + \frac{2}{\sqrt{N}} \sum_{i=0}^{M-1} (-1)^{m.k_{i}}|k_{i}\rangle \nonumber\\
  &= (1 - \frac{4M}{N})|\psi_{m-k}\rangle + ((1 - \frac{4M}{N})\frac{1}{\sqrt{N}} + \frac{2}{\sqrt{N}})\sum_{i=0}^{M-1} (-1)^{m.k_{i}}|k_{i}\rangle 
\end{align} 
The probability of consolidated $M$ phase inverted states is then
$P_{k} = M((1 - \frac{4M}{N}).\frac{1}{\sqrt{N}} + \frac{2}{\sqrt{N}})^{2}$ and for the
$N-M$ non-inverted states $P_{x} = (N - M).((1 - \frac{4M}{N}).\frac{1}{\sqrt{N}})^{2}$.

The ratio of probabilities is thus
\begin{equation}
  \frac{P_{k}}{P_{x}} = \frac{M((1 - \frac{4M}{N}).\frac{1}{\sqrt{N}} + \frac{2}{\sqrt{N}})^{2}}{ (N - M).((1 - \frac{4M}{N}).\frac{1}{\sqrt{N}})^{2}}
\end{equation}
and when $M = \frac{N}{2}$, it reduces to
\begin{equation}
\frac{P_{k}}{P_{x}} = |\frac{\frac{N}{2}}{\frac{N}{2}}| = 1
\end {equation}

With this construction, we can show that again, the relative probability distribuion is not altered by no mens and hence no biasness.
 
The above algorithm is secure against 'chosen plaintext' attack. Even if there will
be single phase inversion, being phase inversion operator as private, key is safe to use
$O(\sqrt{N})$ times in a session. The transmitted message is an equal superposition
state with some co-relation to the key $k$. Without the knowledge of the key
nothing can be inferred about the message ${\it M}$.
To make it completely hardened, approximately half of the total phases of the
Hadamard transformed message state should be inverted.

\subsection{Steps of the algorithm}
The algorithm has the following steps:

\begin{enumerate}
\item Begin with the BB84 (Bennett and Brassard) quantum key distribution (QKD)
  method to establish a shared key $k$ and $d$ between two communicating parties
  (say A and B). Let the key state be $|k\rangle$.
  
\item To each message, $m \in M$, 'A' will apply the Hadamard transform to create
  an equal superposition state $|\psi_{m}\rangle$.
  \[|\psi_{m}\rangle = H^{\otimes n} (|m\rangle) = \frac{1}{\sqrt{N}} \sum_{x=0}^{N-1} (-1)^{m.x}|x\rangle\]

\item Apply the operator $\Lambda_{k}$ to $|\psi_{m}\rangle$ to mark the key state.
 \[ \psi_{c}^{'} = \Lambda_{k} |\psi_{m}\rangle = |\psi_{m}\rangle - 2|k\rangle \]

\item Apply the multiple phase inversion operator $\Upsilon_{k,d}$ to invert
  phases of multiple basis states.
 \[ |\psi_{c}\rangle = \Upsilon_{k,d} |\psi_{c}^{'}\rangle \]

\item Send the resulting encrypted quantum state $|\psi_{c}\rangle$ to 'B' using a quantum channel.
  
\item 'B' will perform the reverse operation to retrieve the message $m$ from $|\psi_{c}\rangle$.
\end{enumerate}

\subsection{Application of proposed cipher}
Some applications of the proposed cipher construction are as follow.

\subsubsection{Authentication}
Anyone can utilize this algorithm to encode and send his/her signature
(public identity) as the message. The intended peer can decode the signature
and verify against the known one. Any tampering of the message would result
in a different signature, i.e. only a entity with the share key in
possession can generate the encoded signature. For authenticated encrypted
message however, total message length will be $2n$ with one part of $n$
bearing the identity for signature verification.

\subsubsection{Quantum Teleportation}
This can be used during quantum teleportation. 'Alice' no longer needs to use
phone or email to communicate 'Bob' her state of operation. Instead, she can
send our cipher to Bob and Bob can 'decrypt' to get the message what Alice had
performed and act accordingly to get the teleported message. Thus, we can
eradicate all classical entities involved in quantum teleportation.

\subsubsection{Rekey}
This can be used to refresh the shared key established by the BB84 QKD. The
newly generated random key $|k_{1}\rangle$ can be communicated to the peer as a
message in our cipher construction using the existing key $|k\rangle$. Aferwards,
$|k_{1}\rangle$ will be the new key and will be used for next set of message
encryption and decryption. In the whole process, BB84 QKD protocol is used only
once. 

\section*{Conclusion}
The proposed quantum cipher is proved mathematically secured against known
attacks (more relevant in the current context 'chosen plaintext attack')
and can be versatile in application. The requirement of a well secure cipher,
namely diffusion and confusion, is satisfied by the Hadamard and multiple phase
inversion transformations respectively. The same idea can be extended and 
similar approach can be used for multi party (multi peers) commuication securely.   


\section*{References}



\end{document}